\documentclass[12pt]{article}
\def\lae{\;^{<}_{\sim} \;} \def\gae{\; ^{>}_{\sim} \;}

\title{\textbf{Supernatural A-term Inflation}}
{\author{\\[1cm]
{\sc \large Chia-Min Lin$^{1}$ and Kingman Cheung$^{2}$}\\
{\sl\small Department of Physics, National Tsing Hua University, Hsinchu, Taiwan 300 }\\
{\sl\small Physics Division, National Center for Theoretical Sciences,
Hsinchu 300, Taiwan}\\
{\sl\small $^3$Division of Quantum Phases \& Devices, School of Physics,
Konkuk university, Seoul 143-701, Korea}
}}

\usepackage[margin=2cm]{geometry}
\usepackage{graphicx,color}
\usepackage{subfigure}
\begin{document}
\maketitle
\begin{abstract}
Following \cite{Lin:2009yt}, we explore the parameter space of the
case when the supersymmetry (SUSY) breaking scale is lower, for
example, in gauge mediated SUSY breaking model. During inflation,
the form of the potential is $V_0$ plus MSSM\footnote{MSSM stands for Minimal Supersymmetric
Standard Model} (or A-term\footnote{It is called A-term inflation when the inflaton field is any direction (gauge or singlet) that generates an A-term.}) inflation. We show that the
model works for a wide range of the potential $V_0$ with the soft SUSY breaking mass $m\sim O(1)$ TeV. The
implication to MSSM (or A-term) inflation is that the flat directions which is
lifted by the non-renormalizable terms described by the
superpotential $W=\lambda_p \phi^{p-1}/M^{p-3}_P$ with $p=4$ and
$p=5$ are also suitable to be an inflaton field for $\lambda_p=O(1)$
provided there is an additional false vacuum term $V_0$ with
appropriate magnitude. The flat directions corresponds to $p=6$ also works for $0 \lae
 V_0/M_P^4 \lae 10^{-40}$.
\end{abstract}
\footnoterule{\small $^1$cmlin@phys.nthu.edu.tw, $^2$cheung@phys.nthu.edu.tw}

\section{Introduction}

Inflation \cite{Starobinsky:1980te, Sato:1980yn, Guth:1980zm} (for
review, \cite{Lyth:1998xn, Lyth:2007qh, Linde:2007fr}) is an vacuum-dominated
epoch in the early Universe when the scale factor grew
exponentially. This scenario is used to set the initial condition for
the hot big bang model and to provide the primordial density perturbation as the
seed of structure formation. In the framework of slow-roll
inflation, the slow-roll parameters are defined by
\begin{equation}
\epsilon \equiv \frac{M_P^2}{2} \left(\frac{V^\prime}{V}\right)^2,
\end{equation}
\begin{equation}
\eta \equiv M_P^2\frac{V^{\prime\prime}}{V},
\end{equation}
where $M_P=2.4\times 10^{18} \mbox{ GeV}$ is the reduced Planck mass.
The spectral index can be expressed in terms of the slow-roll parameters as
\begin{equation}
n_s=1+2\eta-6\epsilon.
\label{n}
\end{equation}
The latest WMAP 5-year result prefers the spectral index around
$n_s=0.96$ \cite{Komatsu:2008hk}.
The spectrum is given by
\begin{equation}
P_R=\frac{1}{12\pi^2M_P^6}\frac{V^3}{V'^2} \;.
\end{equation}
With the slow-roll approximation
the value of the inflaton field $\phi$,
in order to achieve $N$ e-folds inflation, is
\begin{equation}
N=M^{-2}_P\int^{\phi(N)}_{\phi_{end}}\frac{V}{V'}d\phi.
\label{efolds}
\end{equation}
From observation \cite{Komatsu:2008hk} $P^{1/2}_R\simeq 5\times
10^{-5}$ at $N \simeq 60$ (we call this CMB (Cosmic Microwave
Background) normalization).

The scalar potential for the inflaton field $\phi$ of supernatural inflation \cite{Randall:1995dj, Randall:1996ip} (during inflation) is
\begin{equation}
V=V_0+\frac{1}{2}m^2\phi^2
\label{eq1}
\end{equation}
where $V_0 \equiv M_S^4$ is the SUSY breaking scale with $M_S \simeq 10^{11}$ GeV and the second term is the soft mass term with $m \sim O(1)$ TeV in order to address the hierarchy problem of the Standard Model (SM). Supernatural inflation is very attractive but predicts the spectral index $n_s \gae 1$. In \cite{Lin:2009yt}, a model is proposed with the potential includes the A-term and non-renormalizable terms
\begin{equation}
V=V_0+\frac{1}{2}m^2 \phi^2-A\frac{\lambda_p \phi^p}{p M^{p-3}_P}+\lambda^2_p \frac{\phi^{2(p-1)}}{M_p^{2(p-3)}}.
\label{eq2}
\end{equation}
Where $4\leq p \leq9$ for the Minimal Supersymmetric Standard Model (MSSM) \cite{Gherghetta:1995dv}. When $M_S$ is the gravity mediated SUSY breaking scale it is shown that the spectral index can naturally be $n_s=0.96$, because the potential is naturally of a hilltop form (so called hilltop inflation \cite{Boubekeur:2005zm, Kohri:2007gq}
).

The soft mass $m$ is related to $M_S$ via
\begin{eqnarray}
m &\sim& \sqrt{3}CM^2_S/M_P\\
  &=& Cm_{\frac{3}{2}},
\label{eq3}
\end{eqnarray}
where \cite{Dimopoulos:2002kt}
\begin{eqnarray}
\nonumber C &\sim& 1 \;\;\;\;\;\;\;\;\;\mbox{gravity-mediated}\\ \nonumber
C &\gg&  1 \;\;\;\;\;\;\;\;\;\mbox{gauge/gaugino-mediated}\\
C &\sim& 10^{-3} \;\;\;\;\mbox{anomaly-mediated}.
\end{eqnarray}
As we can see from Eq. (\ref{eq3}), for gravity-mediated SUSY
breaking, $M_S \sim 10^{11}$ GeV. For gauge mediated SUSY breaking,
$M_S$ can be as low as $O(1)$ TeV. From Eq. (\ref{eq1}), it is clear
that the original model of supernatural inflation cannot work for
lower $M_S$. However, for the model Eq. (\ref{eq2}), it certainly
can. The reason is even if we have $V_0=0$ the potential still can
accommodate for a successful inflation model if inflation occurs near a saddle point where $V^{\prime}=V^{\prime\prime}=0$, and it is called MSSM
inflation or A-term inflation in the literature
\cite{Allahverdi:2006iq, Lyth:2006ec, Bueno Sanchez:2006xk,
Allahverdi:2008zz} . In this paper, we will explore the parameter
space for a successful inflation model from Eq. (\ref{eq2}) near a saddle point with a
wide range of $V_0$ from $M_S=1$ TeV to $M_S=10^{11}$ GeV.

The paper is organized as follows. We present the calculation and
result in Section \ref{2}. Our conclusions are summarized in Section
\ref{3}.
\section{Calculation}
\label{2} As in the case of MSSM inflation, the potential (Eq.
(\ref{eq2})) has a saddle point where
$V^{\prime}=V^{\prime\prime}=0$ if $m$ and $A$ are related via

\begin{equation}
m^2=\frac{A^2}{8(p-1)}
\end{equation}

The saddle point is at $\phi=\phi_0$ where

\begin{equation}
\phi_0=\left(\frac{m M_P^{p-3}}{\lambda_p \sqrt{2p-2}}\right)^{\frac{1}{p-2}}.
\label{eq4}
\end{equation}

The potential at the saddle point is

\begin{equation}
V(\phi_0) \sim m^2 \phi^2_0+V_0
\end{equation}
Near the saddle point, the potential can be described as
\begin{equation}
V=V(\phi_0)+\frac{1}{6}V^{\prime\prime\prime}(\phi_0)(\phi-\phi_0)^3
\end{equation}
where
\begin{equation}
V^{\prime\prime\prime}(\phi_0)\sim \frac{m^2}{\phi_0}
\end{equation}
From Eq. (\ref{efolds}), the number of e-folds is
\begin{equation}
N=-\frac{2}{M_P^2}\left(\frac{V_0\phi_0}{m^2}+\phi_0^3\right)\left(\frac{1}{\phi-\phi_0}\right)
\end{equation}
The slow roll parameters $\epsilon$ and $\eta$ are
\begin{equation}
\epsilon=\frac{2(\frac{V_0\phi_0}{m^2}+\phi^3_0)^2}{N^4 M_P^6}
\end{equation}
\begin{equation}
\eta=\frac{\frac{m^2}{\phi_0}(\phi-\phi_0)}{V_0+m^2\phi_0^2}=-\frac{2}{N}
\end{equation}
Inflation ends when $|\eta| \sim 1$.
The spectral index is (for $N=50$)
\begin{equation}
n_s \simeq 1+2\eta=1-\frac{4}{N}= 0.92
\end{equation}
which is the same as A-term inflation. The spectrum is give by

\begin{equation}
P_R=\frac{V_0+m^2\phi^2_0}{12\pi^2}\frac{N^4 M_P^2}{4(\frac{V_0\phi_0}{m^2}+\phi_0^3)^2}
\end{equation}

\begin{figure}[htbp]
\begin{center}
\includegraphics[width=0.45\textwidth]{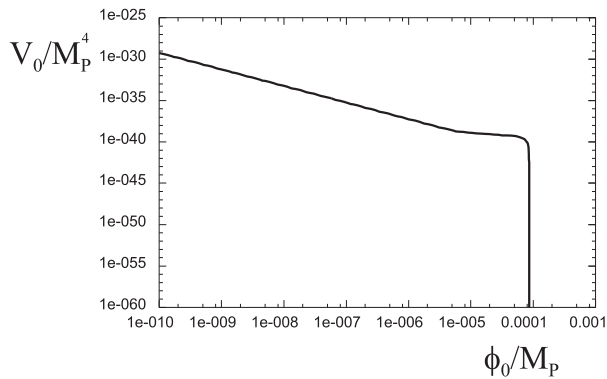}
\caption{Relation between $V_0$ and $\phi_0$. Here we assume $m\sim 1$ TeV.}
\label{result}
\end{center}
\end{figure}

After impose CMB normalization ($P_R=(5 \times 10^{-5})^2$) for
$m=O(1)$ TeV, we obtain the relation between $V_0$ and $\phi_0$. We
show the result on Fig. (\ref{result}).

As we can see from Fig. (\ref{result}), for $10^{-60} \lae
 V_0/M_P^4 \lae 10^{-40}$, the predictions is just the same as
 MSSM (or A-term) inflation. Actually the lower bound can go down to zero, and recovers MSSM (or A-term) inflation. For $10^{-40} \lae V_0/M_P^4 \lae 10^{-30}$, the
 existence of $V_0$ allows also $p=4$ and $p=5$ to fit CMB normalization
 with $\lambda_p=1$. For $p=4$, from Eq. (\ref{eq4}), we have $\phi_0
\simeq 10^{-8}M_P$, corresponds to $V_0/M_P^4 \sim 10^{-35}$.
Similarly, $p=5$ corresponds to $V_0/M_P^4 \sim 10^{-40}$.

\section{Conclusion}
\label{3}

In this paper, we generalize the model of \cite{Lin:2009yt} by considering smaller $V_0$. Our model is similar to hybrid inflation because the additional $V_0$ which eventually should gracefully exist by a waterfall field. However, it is different from the usual hybrid inflation because inflation occurs near a saddle point. The result is for $\lambda_p \sim 1$ and $m\sim O(1)$ TeV, the $p=6$ flat direction can works with $0 \lae V_0/M_P^4 \lae 10^{-40}$. The $p=4$ and $p=5$ flat directions can also play the role of an inflaton field with appropriate $V_0$.

\section*{Acknowledgement}
This work was supported in part by the
NSC under grant No. NSC 96-2628-M-007-002-MY3, by the NCTS, and by the
Boost Program of NTHU. This work is also supported in parts by the WCU
program through the KOSEF funded by the MEST (R31-2008-000-10057-0). 
We are grateful to David Lyth and Kazunori Kohri for discussions.

\end{document}